\documentclass[aps,prl,superscriptaddress,twocolumn,longbibliography]{revtex4-1}
\usepackage{bbm}
\usepackage{graphicx}
\usepackage{dcolumn}
\usepackage{bm}
\usepackage{subfigure}
\usepackage{amsmath}
\usepackage{feynmf}
\usepackage{hyperref}
\usepackage{attachfile}
\usepackage{times}
\usepackage{multirow}
\usepackage[T1]{fontenc}
\DeclareUnicodeCharacter{2212}{-}

\begin{document}

\title{Manipulation of topological phase transitions and the mechanism of magnetic interactions in Eu-based Zintl-phase materials}

\author{Bo-Xuan Li}    
\affiliation{Beijing National Laboratory for Condensed Matter Physics and Institute of Physics, Chinese Academy of Sciences, Beijing 100190, China}
\affiliation{University of Chinese Academy of Sciences, Beijing 100049, China}

\author{Ziyin Song}
\affiliation{Beijing National Laboratory for Condensed Matter Physics and Institute of Physics, Chinese Academy of Sciences, Beijing 100190, China}
\affiliation{University of Chinese Academy of Sciences, Beijing 100049, China}

\author{Zhong Fang}
\affiliation{Beijing National Laboratory for Condensed Matter Physics and Institute of Physics, Chinese Academy of Sciences, Beijing 100190, China}
\affiliation{University of Chinese Academy of Sciences, Beijing 100049, China}

\author{Zhijun Wang}
\affiliation{Beijing National Laboratory for Condensed Matter Physics and Institute of Physics, Chinese Academy of Sciences, Beijing 100190, China}
\affiliation{University of Chinese Academy of Sciences, Beijing 100049, China}

\author{Hongming Weng}
\email{hmweng@iphy.ac.cn}
\affiliation{Beijing National Laboratory for Condensed Matter Physics and Institute of Physics, Chinese Academy of Sciences, Beijing 100190, China}
\affiliation{University of Chinese Academy of Sciences, Beijing 100049, China}
\affiliation{Songshan Lake Materials Laboratory, Dongguan 523808, China}

\date{\today}

\begin{abstract}
Various topological phases, including topological insulators, topological semimetals, and topological superconductors, along with the controllable topological phase transitions, have attracted considerable attention due to their promising applications in spintronics and quantum computing.
In this work, we propose two distinct methods for manipulating topological phase transitions in magnetic materials. 
First, by varying the strength of electron correlation effects, we induce a series of topological state transitions within the EuM$_2$X$_2$ (M = Zn, Cd; X = P, As, Sb) family of Zintl materials, including magnetic topological crystalline insulators (TCIs) and magnetic Dirac semimetals. Our findings indicate that strong electron correlation effects tend to influence the emergence of topological phases.
Second, by reducing the electronegativity of the pnictogen X (from P to As and Sb), we observe a similar transition from trivial insulator to magnetic Dirac semimetal or magnetic TCI. This suggests that weaker electronegativity favors the emergence of topological phases.
Furthermore, we establish a Heisenberg model to describe the magnetic interactions of the EuM$_2$X$_2$ system, based on which we perform Monte Carlo simulations of specific heat and magnetic susceptibility, yielding Néel temperatures that perfectly match experimental data. This suggests that the local magnetic moment framework provides an accurate description of the magnetization behavior in this family of materials.
This work provides the potential for the experimental manipulation of topological phase transitions and their possible applications, while also enhancing the understanding of the magnetic interactions within the EuM$_2$X$_2$ system and offering a theoretical foundation for future applications in magnetism.

\end{abstract}  

\maketitle

\section{I. Introduction}
In recent decades, topological phases have drawn significant attention from condensed matter physicists due to their potential for dissipationless transport and their applications in quantum computation and spintronics \cite{Bi2Se3_zhangHJ2009,inversion_Fu2007,QSH_kane2005,IQHE_klitzing1980,MnBi2Te4_liJH2019_2,QAH_Haldane1988,weyl_WengHW2015,weyl_WengHW2015,Dirac_Young2012,Na3Bi_WangZJ2012,Cd3As2_WangZJ2013,Dirac_ZhangSC2016,Dirac_GangXu2018,PbSnSe_Story2012,TCI_Ando2015,TCI_Fu_2011,SnTe_Ando2012,SnTe_Fu2012}. 
As a result, identifying materials that host topological states and developing methods for controlling these states have become pivotal challenges in condensed matter physics. For materials with a band gap, the topological classification of the material and the corresponding edge states can generally be determined through the calculation of the topological invariants of the occupied states. A topological phase transition is typically accompanied by the closure of the energy gap. Therefore, the capability to control the process of gap opening and closing enables the manipulation of topological states.
Previous studies have explored various approaches to induce and control topological phase transitions, including the application of strain\cite{Lars2013}, external electric fields\cite{You2021}, modification of magnetic configurations\cite{EuCd2As2_MaJZ2020}, adjustment of spin-orbit coupling strength\cite{LQZhou2020,LQZhou2022} and the tuning of interaction effects\cite{XGWan2011iridates,Mohsen2020,WZhang2022}. In this work, we simulate the effect of interaction strength on the topological phase transition by varying the Hubbard-\( U \) value, thereby enabling the control of the topological properties of the compound.

To analyze the process of topological phase transitions, the first step is to identify the topological phase of a given state. Due to the significant impact of symmetry on topological states, magnetic space groups (MSGs) are widely used in the study of topological phases in materials. By calculating the irreducible representations of the energy bands and employing compatibility relations\cite{irvsp,mom2msg}, we can determine band degeneracies and identify band inversions.  Additionally, symmetry indicator (SI) theory allows us to diagnose the topological properties of the system by calculating the eigenvalues of specific symmetry operations at high-symmetry points of the occupied states for a magnetic group\cite{TQC_Bernevig2017,SI_Hoi2017,SI_MSG_Hoi2018,SI_SongZD2018,SI_nsoc_SongZD2018,SI_TCI_Hoi_2018,SI_MSG_JiangY2022,catalogue_ZhangTT_2019}.

Zintl-phase materials consist of metallic cations and anionic frameworks, where the metal cation donates electrons to the anionic framework, forming a strongly covalent anionic structure \cite{Zintl_PeiYZ_2017,Zintl_RenZF2017,Zintl_ZhangQ_2019_2}. Due to their unique structure, Zintl phases exhibit both metallic and covalent characteristics, often leading to small band gaps. This feature allows spin-orbit coupling (SOC) to play a potentially significant role in determining the topological properties of these materials. Consequently, various topological states have been realized in Zintl phases, including topological insulators and Dirac semimetals, fueling substantial interest in their study \cite{Zintl_Damian2024,Zintl_Vanderbilt2022,Zintl_YFXu2019,Zintl_JYYao2025}. For instance, in recent years, numerous experimental and theoretical studies have focused on the magnetic configurations and Néel temperature of EuCd$_2$As$_2$, a compound generally believed to exhibit an A-type antiferromagnetic configuration with magnetic moments oriented along the in-plane \( a \)-axis (AFM-Aa) and a Néel temperature of approximately 9.5 K \cite{Dirac_GangXu2018,EuCd2As2_Inga2011,EuCd2AS2_Maitra2018,EuCd2As2_MaJZ_2019_2,EuCd2As2_MaJZ2020,EuCd2As2_Rahn2018,EuCd2As2_Soh2019,EuCd2As2_WangLL2019,EuCd2As2_WangNL2016}. Above the Néel temperature, a competition between ferromagnetic and antiferromagnetic interactions is observed \cite{EuCd2As2_WangNL2016,EuCd2As2_Yu2012}. When the magnetic moments are aligned in-plane, the compound behaves as a \( Z_2 \) TCI, whereas when the moments align along the \( c \)-axis, the compound exhibits Dirac semimetal behavior. This duality gives rise to the possibility of realizing multiple topological phases within a single compound \cite{EuCd2As2_MaJZ2020}.

Moreover, recent studies have demonstrated that the magnetic configuration of EuCd$_2$As$_2$ can be tuned via crystal growth techniques \cite{EuCd2As2_Manip_Mag_Jo_2020}. In multilayer EuCd$_2$As$_2$, the quantum anomalous Hall (QAH) effect has been realized \cite{Hubbard_NiuCW2019}. Additionally, EuM$_2$X$_2$ compounds have exhibited other intriguing properties, such as colossal magnetoresistance (CMR) \cite{Magnetoresistance_Rahman2024,Magnetoresistance_WangZC2021,Magnetoresistance_ZhangHL2023}, controllable anomalous Hall conductivity \cite{EUM2X2_Mizuki2022}, and thermoelectric effects \cite{EuM2X2_Takagiwa2017}. Clearly, Eu-based Zintl phases present complex and controllable magnetic interactions alongside diverse topological phases, providing opportunities for realizing various topological states in experiments. Furthermore, these materials exhibit numerous potentially exploitable physical properties, warranting further investigation. 

In this study, we investigate the influence of electronic correlations and electronegativity on the topological states of Eu-based Zintl phases, as well as the mechanisms driving magnetic interactions within the EuM$_2$X$_2$ family of materials. Our findings reveal that strong electron correlation effects tend to influence the emergence of topological phases in these materials while the reduction in electronegativity promotes the appearance of topological phases. Moreover, we propose that the local magnetic moment framework offers an accurate description of the magnetization behavior in this family of materials.

\section{II. Method}
We performed first-principles calculations using the Vienna Ab-initio Simulation Package (VASP) \cite{VASP_DFT_Kohn1965,VASP_1996}, with plane-wave basis sets and the Projector Augmented-Wave (PAW) method \cite{VASP_PAW1994} to compute the electronic band structure. The exchange-correlation functional was described using the Perdew, Burke, and Ernzerhof (PBE) form under the Generalized Gradient Approximation (GGA) \cite{VASP_GGA_PBE_1_1996,VASP_GGA_PBE_2_1996}. To simulate the strong correlation effects of Eu-f electrons in the antiferromagnetic phase, we used the GGA+Hubbard-$U$ (GGA+$U$) approach \cite{VASP_Hubbard_U1998}. To simulate the interaction-driven topological phase transitions, we varied the Hubbard-$U$ parameter from 1 eV to 7 eV. The plane-wave cutoff energy was set to 500 eV, and the convergence criterion for total energy was set to $10^{-7}$ eV. For various magnetic configurations, we employed a Monkhorst-Pack k-point sampling scheme \cite{VASP_Monkhorst_1976} with an $11\times11\times5$ k-point grid in the Brillouin zone. The effects of SOC were considered in all calculations. Irreducible representations and symmetry indicators were calculated using Irvsp\cite{irvsp}.

We utilized the software {\small WANNIER90} \cite{wannier90_2014} to construct a tight-binding Hamiltonian model based on maximally localized Wannier functions (MLWFs). Taking EuZn$_2$As$_2$ with the AFM-Aa magnetic configuration as an example, the local Wannier orbital basis included the $p$ orbitals of As, the $s$ and $p$ orbitals of Zn, and the $s$ and $d$ orbitals of Eu. The inner energy window spanned from 6 eV below to 6 eV above the Fermi level, while the outer energy window extended from 6 eV below to 15 eV above the Fermi level. We use the \textit{postw90} module in {\small WANNIER90} \cite{wannier90_2014} to implement the optical conductivity calculation where a $150 \times 150 \times 150$ $k$-mesh is employed. For simulations of temperature-dependent properties, including magnetic susceptibility and specific heat, we employed the Espins package \cite{espins_2022}, which relies on standard Monte Carlo methods. To calculate the surface states of the TCI, we applied the software WannierTools \cite{wanniertools_2018}, which implements the iterative Green’s function method. A principal layer of two and an energy interval of 2000 slices were utilized in the calculation.

\section{III. Results and Discussion}
\subsection{A. Manipulation of topological phase transition}

\begin{figure}
    \centering
    \includegraphics[width=1.0\linewidth]{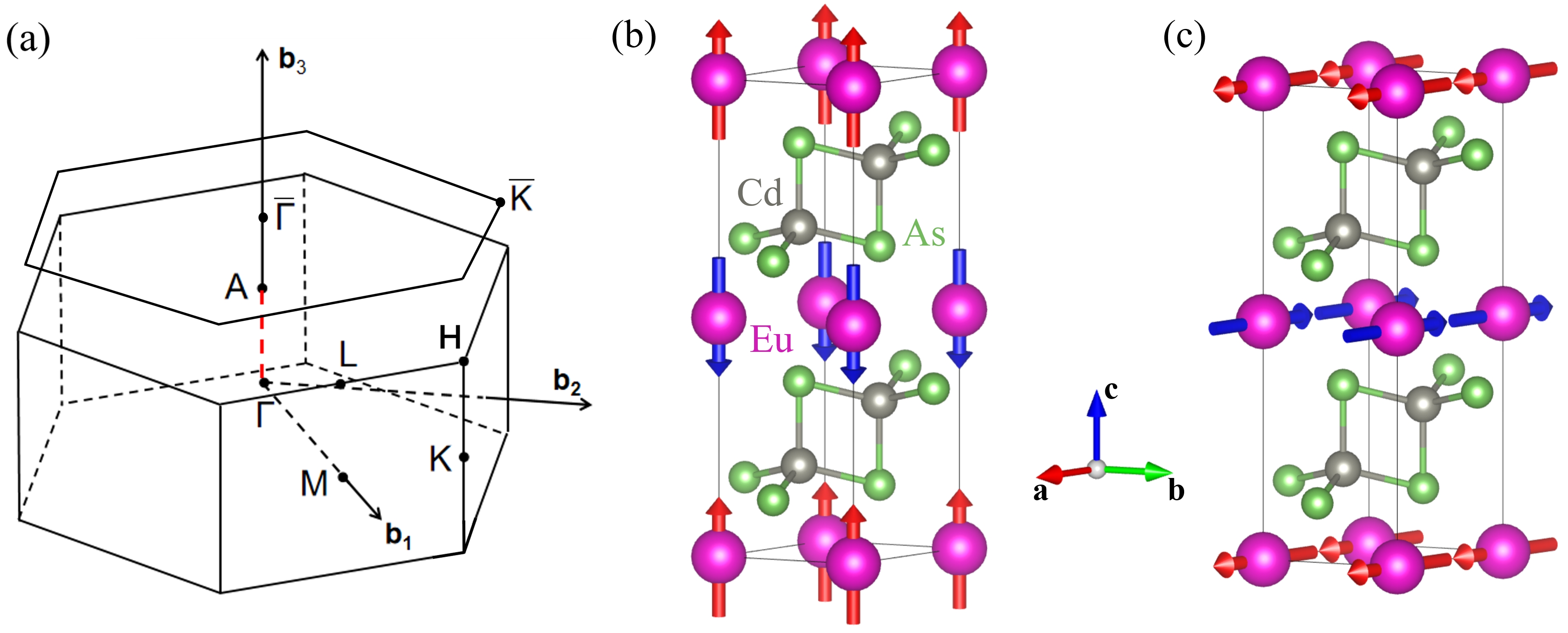}
    \caption{(a) Schematic of the Brillouin Zone of EuCd$_2$As$_2$ (b) Magnetic structure of A-type AFM EuCd$_2$As$_2$, with magnetic moment along \textit{c} axis (AFM-Ac) (c) Magnetic structure of A-type AFM EuCd$_2$As$_2$, with magnetic moment along \textit{a} axis (AFM-Aa)}
    \label{fig1}
\end{figure}

To explore the topological properties of the EuM$_2$X$_2$ family materials, we varied the Hubbard-$U$ parameter from 1 eV to 7 eV to better capture the correlated interactions of the $4f$ electrons. It is well known that the $4f$ electrons in rare earth elements, especially lanthanides, are more localized than the outer $5p$ and $6s$ electrons. Thus, the correlation effects between electrons cannot be neglected, necessitating the inclusion of the Hubbard-$U$ correction to accurately describe rare earth elements with $4f$ orbitals.

Here, we take EuCd$_2$As$_2$ as an example. Experimental data indicate that its magnetic configuration of ground state is AFM-Aa \cite{EuCd2As2_Rahn2018}, as shown in Fig.\ref{fig1}(c). Our calculations reveal that the energy of the magnetic configuration AFM-Ac is only 0.16 meV per magnetic unit cell, slightly higher than that of the magnetic configuration AFM-Aa. For comparison, we calculated the topological properties of the EuM$_2$X$_2$ series for both magnetic configurations with different $U$ values.

The EuM$_2$X$_2$ family materials belong to space group P$\bar{3}$m1 (No. 164). Fig.\ref{fig1}(a) shows the corresponding Brillouin zone and the high-symmetry points. When the magnetic moments of the Eu atoms align along the \emph{c}-axis, forming the AFM-Ac magnetic configuration shown in Fig.\ref{fig1}(b), the corresponding magnetic space group is P$_{\mathrm{c}}\bar{3}$c1 (No. 165.96)\cite{mom2msg}. When the magnetic moments lie within the \emph{ab}-plane, forming the AFM-Aa magneitc configuration shown in Fig.\ref{fig1}(c), the corresponding magnetic space group is C$_{\mathrm{c}}$2/m (No. 12.63). Notice that, both the MSGs contain the inversion symmetry operation $P$ combined with nonsymmorphic time-reversal symmetry $T*\tau$ operation, where $\tau$ represents a fractional translation along a lattice vector. As a result, the bands are doubly degenerate, protected by this symmetry.

\begin{figure*}
    \centering
    \includegraphics[width=1\linewidth]{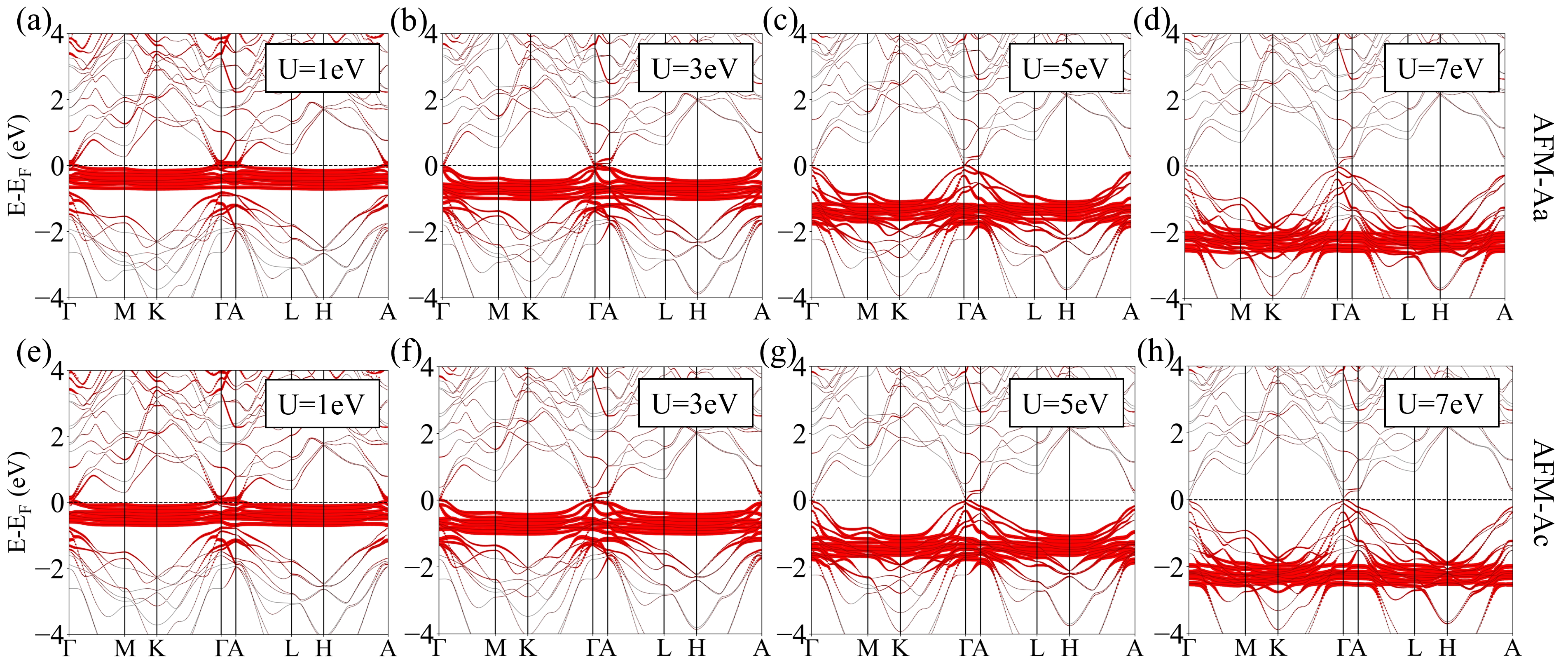}
\caption{Band Structure and Eu-$4f$ Orbital Projection of EuCd\textsubscript{2}As\textsubscript{2}. Effective Hubbard-$U$ interactions of 1, 3, 5, and 7 eV were applied in the calculations for panels (a)-(d), corresponding to the AFM-Aa magnetic configuration, and for panels (e)-(h), corresponding to the AFM-Ac configuration.}
    \label{fig2}
\end{figure*}

Increasing $U$ gradually simulates the process of increasing electron correlation interactions in the material. This is analogous to applying pressure to the material\cite{pressure_BiWL2023}, resulting in the atoms moving closer, thereby enhancing electron correlation effects. Thus, tuning the Hubbard-$U$ value to drive phase transitions has practical significance. Fig.\ref{fig2} illustrates the evolution of the EuCd$_2$As$_2$ band structure for both configurations as Hubbard-$U$ increases. The most notable effect of increasing $U$ on the band structure is the gradual downward shift of the occupied $4f$ electron states. Orbital projection analysis reveals that the primary orbital components near the Fermi level are the \emph{p} orbitals of As and the \emph{s} orbitals of Cd (see Supplemental Material Fig.S1).

\begin{figure}
    \centering
    \includegraphics[width=1\linewidth]{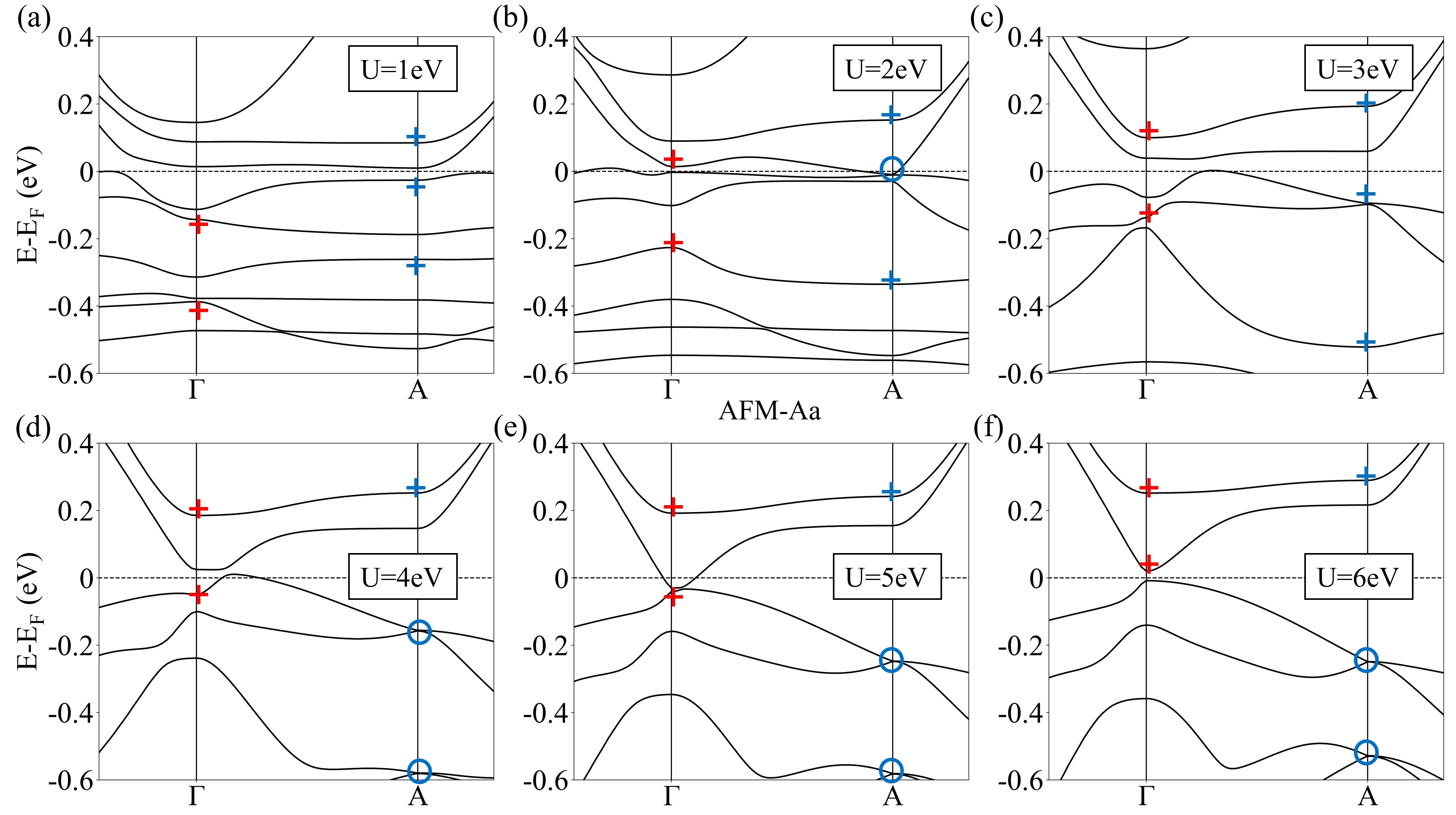}
    \caption{Band structure of EuCd\textsubscript{2}As\textsubscript{2} with the AFM-Aa magnetic configuration along the $\Gamma$-A path. Panels (a) to (f) depict the evolution of band structures for Hubbard-$U$ values from 1 to 6 eV. The band structure for $U$ = 7 eV is omitted as it closely resembles that of $U$ = 6 eV. Bands at the $\Gamma$ point with positive parity eigenvalues are marked with red plus signs, while those at the A point with eigenvalue $i$ of the $C_2$ operation are denoted by blue plus signs. Degenerate bands with both $i$ and $-i$ eigenvalues of the $C_2$ operation are represented by blue circles.}
    \label{Fig3}
\end{figure}

Topological quantum chemistry and symmetry indicators have emerged in recent years as powerful tools for analyzing the topological classification of compounds. These methods are highly useful in facilitating high-throughput searches for topological materials\cite{TQC_Bernevig2017,SI_Hoi2017,SI_MSG_Hoi2018,SI_SongZD2018,SI_nsoc_SongZD2018,SI_TCI_Hoi_2018,SI_MSG_JiangY2022,catalogue_ZhangTT_2019}. In addition to analyzing the relative positions of bands and corresponding band inversions, we also calculated the SI, $Z'_{2p}$, for different $U$ values based on the parity of all occupied states below the Fermi level, as described in \cite{SI_MSG_JiangY2022}:
\begin{equation}
    Z'_{2p} = \sum_{k \in \text{TRIM}} \frac{1}{4} (N_k^- - N_k^+) \mod 2.
\end{equation}
We notice that the presence of a significant gap within the occupied states, so we only need to count the parity information of the states between the gap and the Fermi level. Note that, through the analysis of the representation of MSG for both magnetic configuration, among the eight time-reversal invariant momenta (TRIM) in the Brillouin zone, all band pairs at four points in the $k_{z} = \pi/c$ are energy-degenerate with opposite parity eigenvalues, and thus do not contribute to the $Z_{2}$ index. Therefore, we only need to consider the four points in the $k_{z} = 0$ plane in the following analysis. Notably, in the following discussion, each pair of doubly degenerate bands is assigned a single parity eigenvalue.

For the AFM-Aa magnetic configuration of EuCd$_2$As$_2$, we present the band structure near the $\Gamma$ point in the vicinity of the Fermi level in Fig.\ref{Fig3}, along with the parity eigenvalues of each band at the $\Gamma$ point and the eigenvalues of $C_2$ operation at the A point (where $+i$ is denoted as $+$, and $-i$ as $-$). Since only one representation, $B3+B4$, exists along the $\Gamma$-A path, there is no crossing between the conduction and valence bands, ensuring a gap. Thus, to analyze the topological transition, we focus only on the number of bands below the Fermi level with positive parity at the $\Gamma$ point. We observe that there is a change in the number of bands with positive parity below the Fermi level at the $\Gamma$ point as $U$ increases from 1 eV to 2 eV. This corresponds to a topological phase transition from a trivial insulator to a Z$_2$ TCI. As $U$ increases to 3, 4, and 5 eV, the number of occupied states with positive parity at the $\Gamma$ point remains unchanged. However, when $U$ increases from 5 eV to 6 eV, a second change in the number of occupied states with positive parity at the $\Gamma$ point occurs, indicating a phase transition from a TCI back to a trivial insulator. The specific parity eigenvalues at each point are shown in Table\ref{table_parity}.

\begin{table}[h]
\centering
\caption{The number of occupied bands with positive parity eigenvalues among the four high symmetry $k$ points at $k_z = 0$ within the Brillouin zone for the AFM-Ac and AFM-Aa magnetic configurations of EuCd$_2$As$_2$, below the Fermi energy, along with their corresponding topological classifications at varying values of $U$.}
\label{table_parity}
\begin{tabular}{ccccccccc}
\hline\hline
 & \multicolumn{3}{c}{AFM-Ac} & & \multicolumn{4}{c}{AFM-Aa} \\
\cline{2-4} \cline{6-9}
$U$ (eV)& $\Gamma$ & 3M & CLASS & & $\Gamma$ & Y & 2V & CLASS\\
\hline
1 & 8 & 6 & Trivial & & 8 & 6 & 6 & Trivial \\
2 & 7 & 6 & TCI & & 7 & 6 & 6 & TCI \\
3 & 7 & 6 & TCI & & 7 & 6 & 6 & TCI \\
4 & 7 & 6 & Dirac & & 7 & 6 & 6 & TCI \\
5 & 6 & 6 & Trivial & & 7 & 6 & 6 & TCI \\
6 & 6 & 6 & Trivial & & 6 & 6 & 6 & Trivial \\
7 & 6 & 6 & Trivial & & 6 & 6 & 6 & Trivial \\
\hline\hline
\end{tabular}
\end{table}

\begin{figure}
    \centering
    \includegraphics[width=1\linewidth]{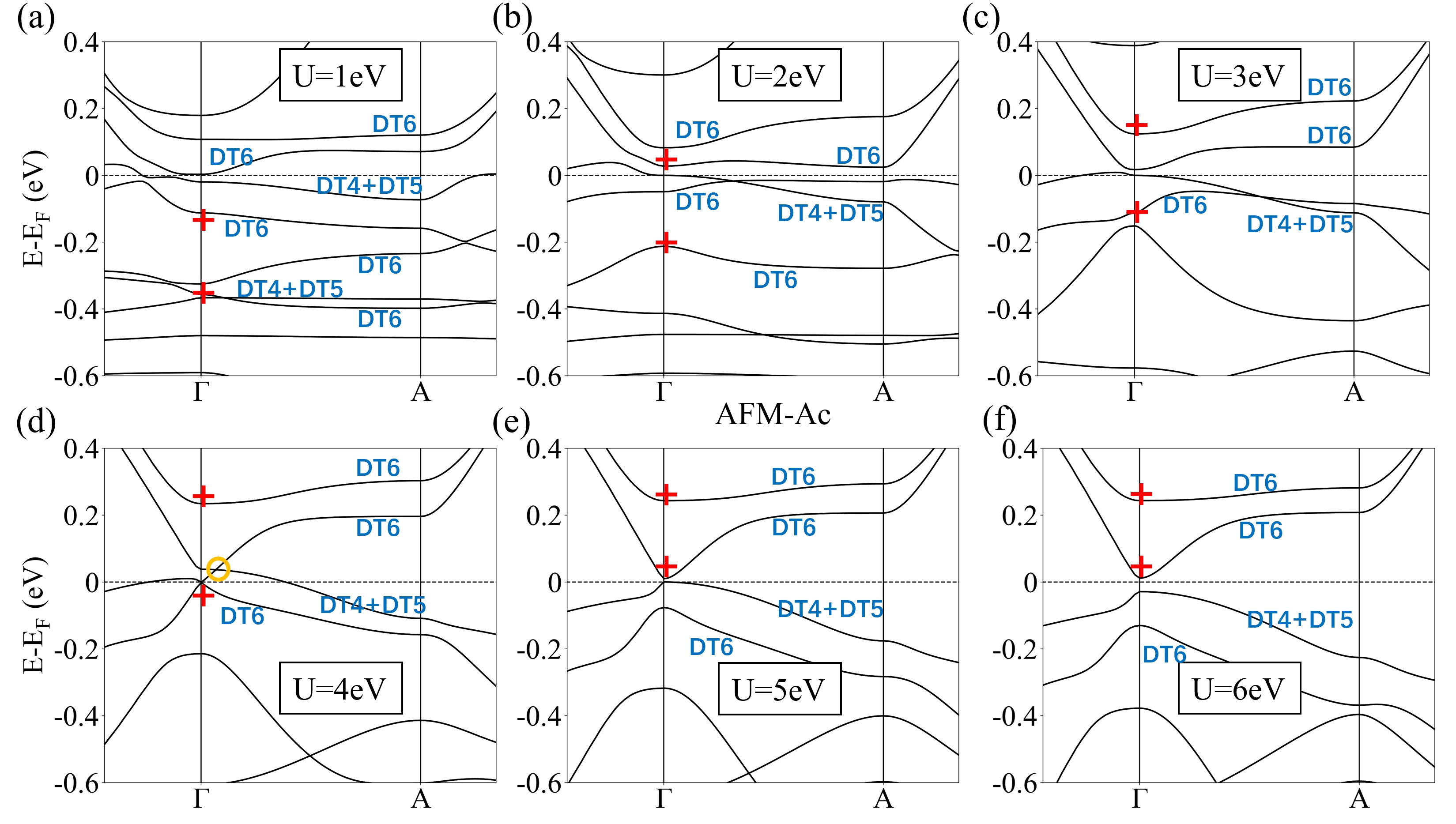}
    \caption{Band Structure along the $\Gamma$-A path for EuCd\textsubscript{2}As\textsubscript{2} with the AFM-Ac magnetic configuration. (a)-(f) correspond to Hubbard-$U$ values ranging from 1 to 6 eV, with the band structure for $U$ = 7 eV being nearly identical to that for $U$ = 6 eV and thus omitted here. The bands at the $\Gamma$ point with positive parity eigenvalues are indicated by red plus signs. Additionally, (b)-(d) show the representations along the $\Gamma$-A path for the bands that cross near the Fermi level.}
    \label{Fig4}
\end{figure}

In Fig.\ref{Fig4}, we present the band structure near the $\Gamma$ point for EuCd$_2$As$_2$ with magnetic configuration AFM-Ac, along with the bands with positive parity eigenvalues near the Fermi level at the $\Gamma$ point, and the corresponding representations of the bands along the high-symmetry $\Gamma$-A path. Similarly, as $U$ increases from 1 eV to 7 eV, the number of bands below the Fermi level with positive parity at the $\Gamma$ point decreases from 8 to 7, and then to 6. This indicates that two band inversions occur as $U$ rises. Consistently, the calculations of SI show that at $U$ = 1 eV, the compound is a trivial insulator; at $U$ = 2 and 3 eV, it is a Z$_2$ TCI; at $U$ = 4 eV, becomes a Dirac semimetal; and at $U$ = 5, 6, and 7 eV, the compound returns to a trivial insulator. The specific parity transformations are listed in Table\ref{table_parity}.
In the following, we present a detailed description of the band inversion process as the Hubbard-$U$ value increases.

{\it As $U$ increases from 1 eV to 2 eV:} The second band below the Fermi level at $U$ = 1 eV (labeled as DT6 along the $\Gamma$-A path, with a positive parity eigenvalue at the $\Gamma$ point) exchanges order with the two bands above it, resulting in the first band above the Fermi level at $U$ = 2 eV having a positive parity eigenvalue at the $\Gamma$ point. The first band below the Fermi level at $U$ = 1 eV moves up and intersects with the first band above it. Due to the different representations of the two bands, no gap opens at the intersection, as shown in Fig.\ref{Fig4}(b). The fourth band below the Fermi level at $U$ = 1 eV (labeled as DT6 along the $\Gamma$-A path, with a positive parity eigenvalue at the $\Gamma$ point) undergoes a band inversion at the $\Gamma$ point with the band above it. Since both bands along the $\Gamma$-A path are labeled as DT6, a gap opens at the band crossing, as shown by the third band below the Fermi level in Fig.\ref{Fig4}(b).

{\it As $U$ increases from 2 eV to 3 eV:} The first and second bands above the Fermi level at $U$ = 2 eV undergo a band inversion at the $\Gamma$ point. Since both bands are labeled as DT6, no crossing appears along the $\Gamma$-A path. The third and the second band below the Fermi level undergo a band inversion at the $\Gamma$ point. Similarly, no crossing appears along the $\Gamma$-A path since both bands share the same DT6 representation.

{\it As $U$ increases from 3 eV to 4 eV:} The first band above the Fermi level at $U$ = 3 eV (with negative parity eigenvalue at the $\Gamma$ point) undergoes a band inversion with the first band below the Fermi level (also with negative parity eigenvalue at the $\Gamma$ point), and the first and second band below the Fermi level approximately intersect at the $\Gamma$ point. Since the bands above and below the Fermi level have different representations at $U$ = 4 eV, a band crossing occurs between the conduction and valence bands during the band inversion. As previously mentioned, the presence of the combined P and $T*\tau$ symmetry ensures that each band is doubly degenerate. Therefore, the crossing point here is a fourfold degenerate point, also known as a Dirac points illustrated by a orange circle in Fig.\ref{Fig4}(d). Meanwhile, the first and second bands below the Fermi level exchange order at A, lifting the degeneracy along the $\Gamma$-A path.

{\it As $U$ increases from 4 eV to 7 eV:} The two energy bands which give rise to the Dirac point at $U$ = 4 eV gradually separate until the Dirac point disappears, resulting in a positive parity eigenvalue for the first energy band above the Fermi surface at the $\Gamma$ point. The band structure of $U$ = 7 eV is almost identical to that of $U$ = 6 eV, so it is omitted from the figure.

\begin{figure*}
    \centering
    \includegraphics[width=1\linewidth]{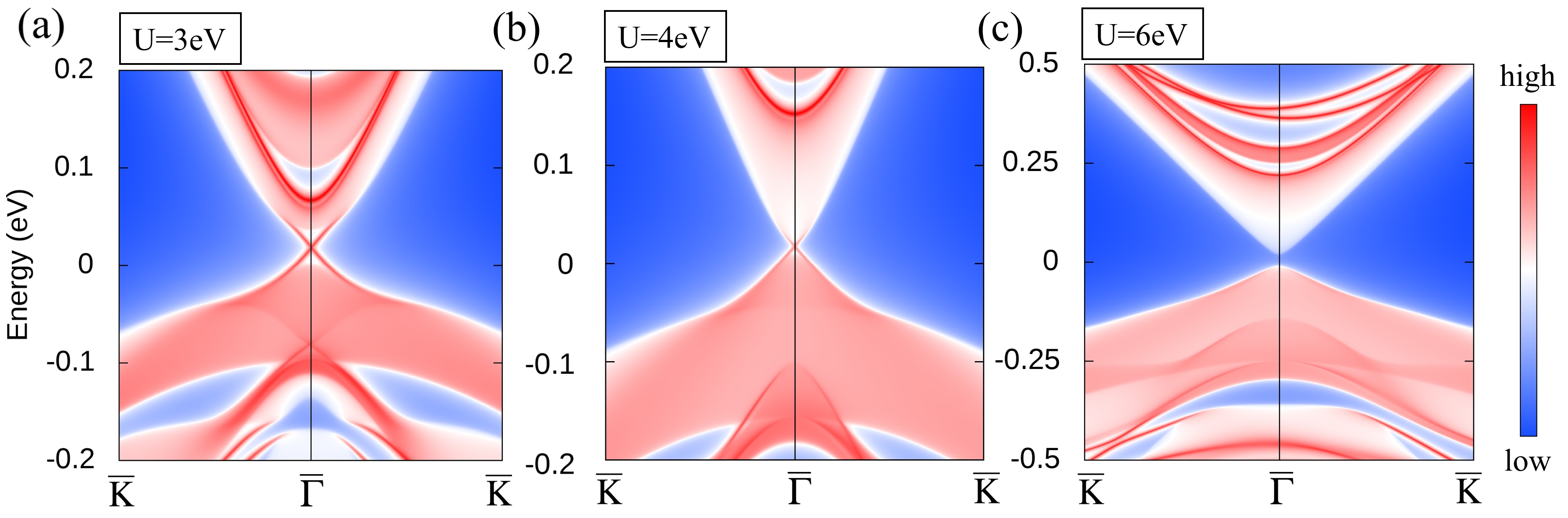}
    \caption{Surface states of EuCd$_2$As$_2$ with AFM-Aa configuration for Hubbard $U$ values of 3,4 and 6 eV}
    \label{figSS}
\end{figure*}

To further investigate the surface states of TCI, we consider the AFM-Aa magnetic configuration of EuCd$_2$As$_2$ as an illustrative example. Among the EuM$_2$X$_2$ series, the AFM-Aa configuration of EuCd$_2$As$_2$ exhibits the most significant band gaps, making it an ideal candidate for visualizing surface states. As shown in \cite{SI_MSG_JiangY2022}, the non-trivial SI, $Z'_{2p}$ signifies the presence of non-trivial topological invariants originating from mirror symmetry and anti-unitary translation symmetry operations. These invariants give rise to surface states in the surface Brillouin zone perpendicular to the mirror plane. A Dirac cone forms along the mirror-invariant high-symmetry line, which corresponds to the intersection of the surface Brillouin zone with the mirror plane in the 3D Brillouin zone. Due to the opposite mirror Chern numbers of two groups of bands with different mirror symmetry eigenvalues, the surface states manifest as a Dirac cone in the surface Brillouin zone rather than as chiral edge states. 

In this primitive unit cell setting, the mirror plane is $M_{100}$. To investigate the surface state, we calculated the band structure along the $\overline{K}-\overline{\Gamma}-\overline{K}$ path in the surface Brillouin zone, explicitly revealing the existence of the surface states. As shown in Table \ref{table_parity} and Fig.\ref{Fig3}, the compound is a TCI for $U$ = 3, 4, and 5 eV, and transitions to a trivial insulator when $U$ = 6 eV. Notably, at $U$ = 3 eV, the band gap is as large as 50 meV, which facilitates the observation of surface states, as shown in Fig.\ref{figSS}(a). For $U$ = 4 eV, the band gap is smaller, yet the surface states remain discernible, as depicted in Fig.\ref{figSS}(b). At $U$ = 6 eV, the compound becomes a trivial insulator. Although a clear gap is observed in Fig.\ref{figSS}(c), no surface states are present, aligning perfectly with the SI analysis described above.

It is worth noting that we observe that the first and second bands below the Fermi level at the A point appear to be 'degenerate' for $U$ values between 4 and 6 eV, as illustrated by a blue circle in Fig.\ref{Fig3}(d)-(f). This 'degeneracy' can also be found between the conduction and valence bands at the A point for $U = 2$ in Fig.\ref{Fig3}(b). Due to the combined symmetry of $T*\tau$ and $P$ in this material ensuring that every band is doubly degenerate, the point where the first and second bands below the Fermi level meet forms a fourfold 'degeneracy'. The MSG of EuCd$_2$As$_2$ in this magnetic configuration is C$_{\mathrm{c}}$2/m (No. 12.63), which has two representations at A, namely A3+A5 and A4+A6, both of which are two-dimensional representations. This is inconsistent with the MSG representations, which only allow for twofold degeneracies. What causes this discrepancy?

Taking the 'degeneracy' between the conduction and valence bands at the A point in Fig.\ref{Fig3}(d) as an example, we find that these seemingly degenerate bands actually exhibit an energy splitting of 0.12 meV at A, indicating that they are not truly degenerate. This confirms the consistency of the MSG representation analysis. However, we find that such approximate fourfold degeneracies, or very small energy splittings at high-symmetry points, are not coincidental in the band structures shown in Fig.\ref{Fig3}. We believe that the theory of spin space group (SSG) \cite{SSG_JiangY2024,SSG_SongZY2024} can explain the occurrence of these approximate fourfold degeneracies. The MSG describes magnetic systems with SOC, whereas SSG is established to describe magnetic systems without SOC. When the effect of SOC is very weak, the energy contribution of SOC can be treated as a perturbation. Consequently, we can use SSG representations to analyze systems with weak SOC. Through calculations, we find that the SSG of this material is 164.2.1.1.L \cite{SSG_SongZY2024}, which has six different representations at A, two of which are four-dimensional, and four of which are two-dimensional. This suggests that if we ignore SOC, the aforementioned approximate fourfold degeneracy would be a true fourfold degeneracy.

\begin{table*}[!ht]
\centering
\caption{Topological classification of AFM-Ac and AFM-Aa magnetic configuration of EuCd$_2$As$_2$ series compounds at varying values of $U$.}
\label{table_classification}
\begin{tabular}{ccccccccc}
\hline\hline
\textbf{Material} & \textbf{Magnetic Configuration} & \textbf{$U$=1 eV} & \textbf{$U$=2 eV} & \textbf{$U$=3 eV} & \textbf{$U$=4 eV} & \textbf{$U$=5 eV} & \textbf{$U$=6 eV} & \textbf{$U$=7 eV} \\
\hline
EuZn$_2$P$_2$ & AFM-Ac & Trivial & Trivial & Trivial & Trivial & Trivial & Trivial & Trivial \\
\hline
EuZn$_2$P$_2$ & AFM-Aa & Trivial & Trivial & Trivial & Trivial & Trivial & Trivial & Trivial \\
\hline
EuZn$_2$As$_2$ & AFM-Ac & TCI & TCI & Dirac & Dirac & Trivial & Trivial & Trivial \\
\hline
EuZn$_2$As$_2$ & AFM-Aa & TCI & TCI & TCI & TCI & Trivial & Trivial & Trivial \\
\hline
EuZn$_2$Sb$_2$ & AFM-Ac & TCI & TCI & Dirac & Dirac & Dirac & Dirac & Dirac \\
\hline
EuZn$_2$Sb$_2$ & AFM-Aa & TCI & TCI & TCI & TCI & TCI & TCI & TCI \\
\hline
EuCd$_2$P$_2$ & AFM-Ac & Trivial & Trivial & Trivial & Trivial & Trivial & Trivial & Trivial \\
\hline
EuCd$_2$P$_2$ & AFM-Aa & Trivial & Trivial & Trivial & Trivial & Trivial & Trivial & Trivial \\
\hline
EuCd$_2$As$_2$ & AFM-Ac & Trivial & TCI & TCI & Dirac & Trivial & Trivial & Trivial \\
\hline
EuCd$_2$As$_2$ & AFM-Aa & Trivial & TCI & TCI & TCI & TCI & Trivial & Trivial \\
\hline
EuCd$_2$Sb$_2$ & AFM-Ac & TCI & TCI & TCI & TCI & Dirac & Dirac & Dirac \\
\hline
EuCd$_2$Sb$_2$ & AFM-Aa & TCI & TCI & TCI & TCI & TCI & TCI & TCI \\
\hline\hline
\end{tabular}
\end{table*}

Furthermore, we have calculated the topological classifications of other materials in the EuM$_2$X$_2$ family materials with varying $U$, as shown in Table\ref{table_classification}. Overall, as the Hubbard-$U$ increases, the increasing correlation interaction among Eu-$4f$ electrons manifests in the band structure as an increasing energy difference between the unoccupied $4f$ orbitals and the 
occupied $4f$ orbitals, as shown in Fig.\ref{fig2}. The energy of the occupied Eu-$4f$ electrons decreases with increasing $U$, which is one driving factor for the band inversion. In summary, by reducing the electronegativity of the pnictogen element X, we enabled the transition from a trivial insulator to a magnetic Dirac semimetal or a magnetic TCI.
Moreover, by analyzing the representations along high-symmetry paths for different magnetic configurations, we find that the symmetry operations in MSG constrain the appearance of possible topological phases, demonstrating that magnetic configurations play a crucial role in their topological classification. We will elaborate on the magnetic interactions of the EuM$_2$X$_2$ family materials in the following sections. 

When comparing the band structures obtained from calculations and experiments for a given material, one specific Hubbard-$U$ value is typically the most suitable. For example, in previous research on Eu-based Zintl-phase materials, Hubbard-$U$ values ranging from 3.1 eV to 6.3 eV have been adopted\cite{EuCd2AS2_Maitra2018,Hubbard_Soh2018,Hubbard_LiBY2022,Hubbard_Mizuki2022,Hubbard_NiuCW2019}. Using EuZn$_2$As$_2$ with magnetic configuration AFM-Aa as an example, by comparing the optical conductivity obtained from calculations with experimental transmission results, we find when $U$ is set to 5 eV, the peaks in optical conductivity match the bottom of transmission spectrum around the wave number of 2700 $cm^{-1}$\cite{ZYLiao2024} very well (see Supplemental Material Fig.S2). Therefore, we use a Hubbard-$U$ of 5 eV in the comparative study of the EuM$_2$X$_2$ series of materials, to control variables effectively.
\begin{figure*}
    \centering
    \includegraphics[width=1\linewidth]{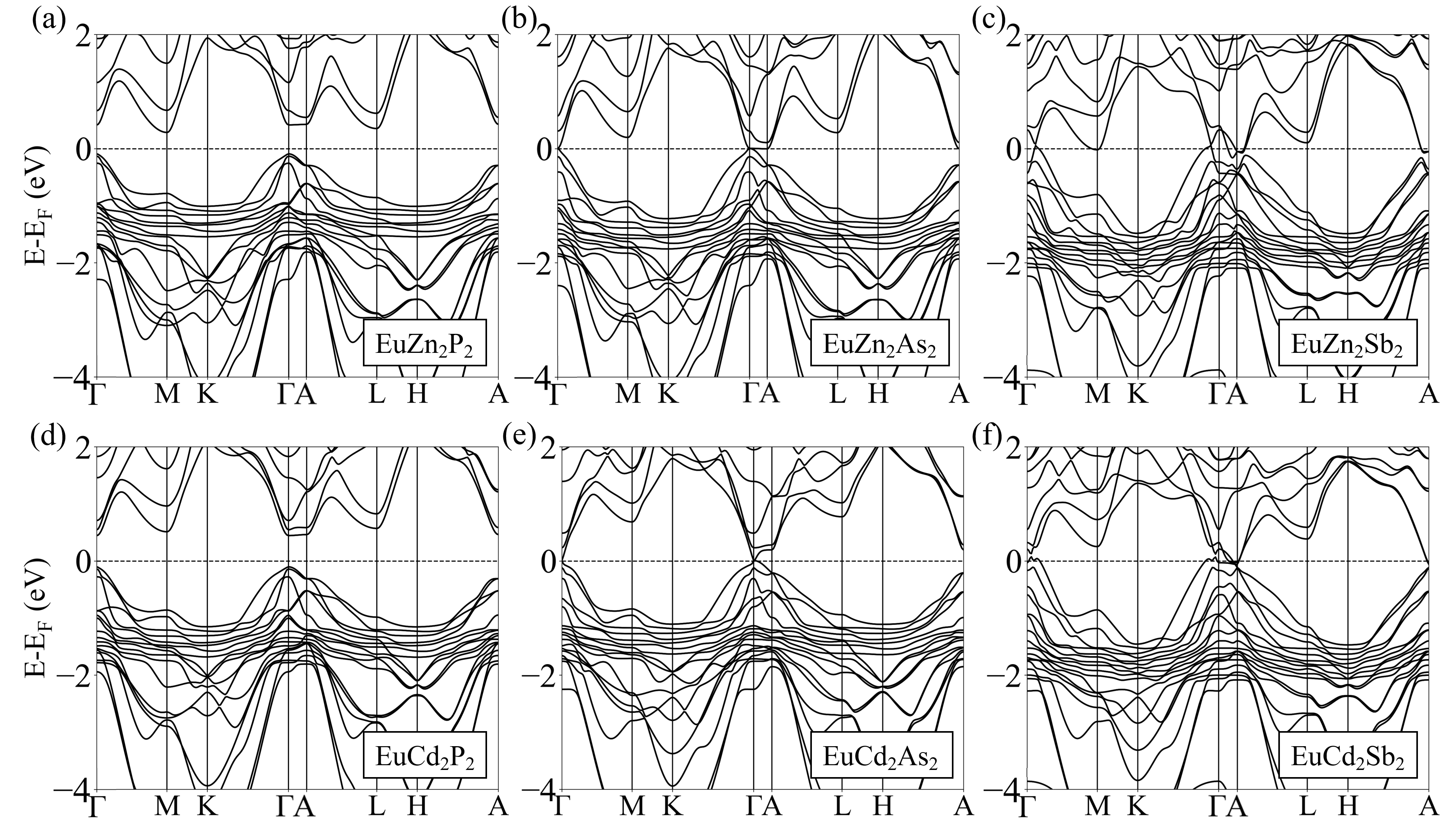}
    \caption{Comparison of band structures for EuM$_2$X$_2$ series materials with AFM-Aa configuration when effective Hubbard interaction of $U$=5 eV is applied.}
    \label{figfamily}
\end{figure*}

In Fig.\ref{figfamily}, we present the band structures of the EuM$_2$X$_2$ series materials with magnetic configuration AFM-Aa at $U$ = 5 eV. From top to bottom and left to right, both M and X atoms are arranged from lighter to heavier. Firstly, comparing the two rows, we observe that the substitution of the atom M has a minimal impact on the band structure. In contrast, a comparison of the three columns from left to right reveals a notable change in the band structure. Specifically, as the atom X becomes heavier, the band gap between the conduction and valence bands decreases gradually. When X is P, a distinct band gap exists between the conduction band and the valence band. As X transitions to As, the band gap nearly vanishes, and when X is Sb, the band structure exhibits a 'negative band gap'. 
We attribute this significant decrease in the band gap from element substitution to differences in their electronegativity. For ionic crystals (e.g. NaCl), the band gap is often large due to the pronounced effect of the transfer of electrons between atoms. In contrast, in covalent compounds, the electron transfer is less pronounced, leading to a smaller band gap at the Fermi level. As the atom X changes from P to As and Sb, its electronegativity reduces. Consequently, the difference in electronegativity between the atom X and the atom Eu or M also reduces. This increase in covalent character of the chemical bonds results in a transition from a positive band gap to a zero band gap and eventually to a negative band gap.

\begin{figure*}
    \centering
    \includegraphics[width=1\linewidth]{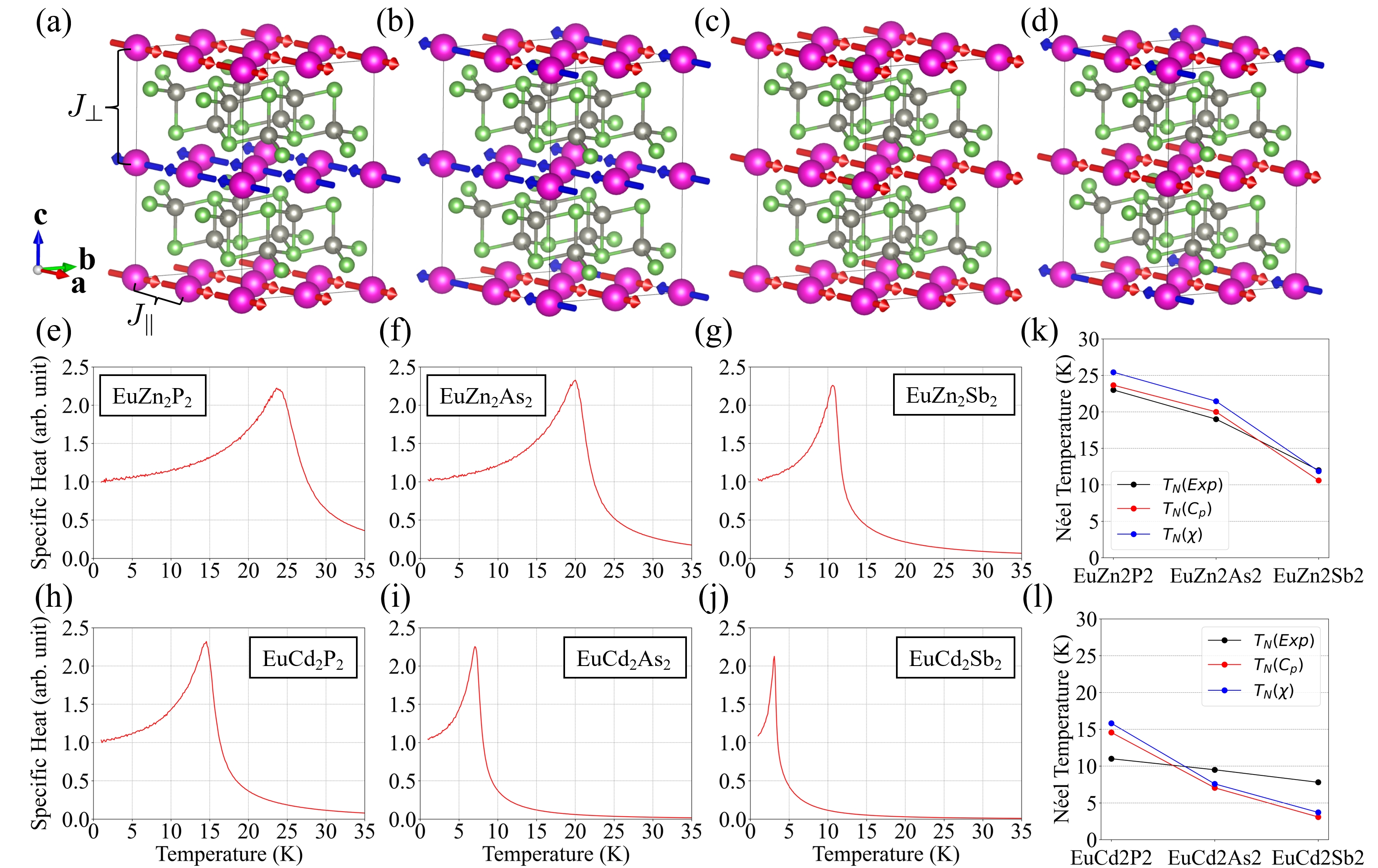}
    \caption{Panels (a)-(d) display the magnetic unit cells of EuM$_2$X$_2$ with distinct magnetic configurations: AFM, AFM$^\prime$, FM, and FM$^\prime$. The temperature dependence of the heat capacity for the EuM$_2$X$_2$ series is presented in panels (e)-(j). Panels (k)-(l) offer a comparative overview of the Néel temperature, juxtaposing theoretical predictions with experimental results for both the EuZn$_2$X$_2$ and EuCd$_2$X$_2$ series. In these figures, black points represents the experimental data, red points corresponds to theoretical calculations based on heat capacity, and blue points signifies those derived from susceptibility data.}
    \label{figneel}
\end{figure*}

\begin{table*}[!ht]
\centering
\caption{Comparison of ground state energies of various magnetic configurations, convergence tolerances, exchange parameters, and N\'eel temperatures for EuM$_2$X$_2$ series compounds. (Note that, 'Exp' represents the experimental data, 'Cp' represents the results derived from heat capacity simulation and '$\chi$' represents the results derived from susceptibity simulation.) }
\label{table_neel}
\begin{tabular}{lcccccc}
\hline\hline
 & EuZn$_2$P$_2$ & EuZn$_2$As$_2$ & EuZn$_2$Sb$_2$ & EuCd$_2$P$_2$ & EuCd$_2$As$_2$ & EuCd$_2$Sb$_2$ \\
\hline
AFM (eV) & -211.297414 & -197.947976 & -181.676900 & -198.008644 & -190.895111 & -177.460445 \\
AFM$^\prime$ (eV) & -211.279861 & -197.933956 & -181.668696 & -197.998412 & -190.890047 & -177.457724 \\
FM (eV) & -211.285872 & -197.930864 & -181.657084 & -197.997200 & -190.890725 & -177.452341 \\
FM$^\prime$ (eV) & -211.274085 & -197.925336 & -181.658776 & -197.992688 & -190.887789 & -177.453747 \\
Tolerance (eV) & $5 \times 10^{-6}$ & $7 \times 10^{-5}$ & $1 \times 10^{-5}$ & $3 \times 10^{-6}$ & $3 \times 10^{-5}$ & $7 \times 10^{-5}$ \\
$J_{\parallel}$ (meV) & 1.22 & 0.812 & 0.271 & 0.614 & 0.334 & 0.0579 \\
$J_{\perp}$ (meV) & -0.721 & -1.07 & -1.24 & -0.715 & -0.274 & -0.506 \\
T$_N$[Exp] (K) & 23 & 19 & 12 & 11 & 9.5 & 7.8 \\
T$_N$[Cp] (K) & 23.65 & 20.00 & 10.60 & 14.57 & 7.05 & 3.09 \\
T$_N$[$\chi$] (K) & 25.43 & 21.46 & 11.86 & 15.82 & 7.58 & 3.71 \\
\hline\hline
\end{tabular}
\end{table*}

\subsection{B. Mechanisms of magnetic interactions in EuM$_2$X$_2$}

In previous studies, the investigation of magnetic configurations and the Néel temperature of the EuM$_2$X$_2$ system has been a prominent topic \cite{EuCd2As2_WangNL2016,EuCd2As2_Rahn2018,EuZn2As2_Blawat2022,EuZn2P2_Berry2022,EuZn2P2_ChenXY2024,EuZn2As2_WangZC2022,EuZn2P2_Krebber2023,Monte_EuZn2P2_Ptok2023,EuCd2Sb2_Soh2018,EuCd2Sb2_Heinrich2024}. Taking EuCd$_2$As$_2$ as an example, previous studies have shown some controversy regarding its ground-state magnetic configuration, however, it is now generally accepted that its ground state adopts the AFM-Aa magnetic configuration. We summarize the Néel temperatures of various materials in the EuM$_2$X$_2$ series, as shown in Table\ref{table_neel} \cite{EuZn2P2_Berry2022,EuCd2As2_Inga2011,EuCd2As2_MaJZ_2019_2,Neel_1,Neel_2,Neel_3,Neel_4,Neel_5,Neel_6,Neel_7}. Observations show that, when the atom M is fixed, the Néel temperature decreases as the atom X transitions from P to As and Sb, indicating a reduction in the stability of the magnetic interactions. To explore the fundamental reason for this trend, we establish a basic Heisenberg model that includes both in-plane and out-of-plane magnetic interactions:

\begin{equation}
H_{\text{eff}} = -\sum_{\langle i,j \rangle} J_{\parallel} \hat{\boldsymbol{S}_i} \cdot \hat{\boldsymbol{S}_j} - \sum_{\langle\langle i^\prime,j^\prime \rangle\rangle} J_{\perp} \hat{\boldsymbol{S}_{i^\prime}} \cdot \hat{\boldsymbol{S}_{j^\prime}}.
\end{equation}

where $J_{\parallel}$ represents the exchange parameter within the $ab$ plane, and $J_{\perp}$ is the exchange parameter along the $c$ direction, as shown in Fig.\ref{figneel}(a). $i,j$ and $i^\prime,j^\prime$ represent the nearest neighbor and next nearest neighbor, respectively. To determine $J_{\parallel}$ and $J_{\perp}$, we calculated the ground state energies for four magnetic configurations\cite{HJXiang2011,HJXiang2013,Monte_EuZn2P2_Ptok2023} : (a) the antiferromagnetic configuration (AFM), (b) the AFM configuration with the spin of a randomly selected atom reversed (AFM$^\prime$), (c) the ferromagnetic configuration (FM), and (d) the FM configuration with the spin of a randomly selected atom reversed (FM$^\prime$) as shown in Fig.\ref{figneel}.

Notably, we solve for three variables namely: $J_{\parallel}$, $J_{\perp}$ and the reference energy $E_0$ using four equations, making it an overdetermined system. Thus, we use the variables derived from first three configurations to verify the ground-state energy of the fourth configuration. The ground-state energy of the fourth configuration calculated in this way, shows negligible deviation from the value obtained through first-principles calculations, indicating that the overdetermined system of equations is self-consistent. To clarify, we have also listed the self-consistent tolerance for each compound in Table\ref{table_neel}. As shown in Table\ref{table_neel}, we notice that the exchange parameter within the $ab$ plane $J_{\parallel}$, is considerably small
when X = Sb, we attribute this to the distance between Eu atoms in the $ab$ plane is 4.494 $\AA$ and 4.698 $\AA$ in EuZn$_2$Sb$_2$ and EuCd$_2$Sb$_2$\cite{Neel_3,Neel_7,Neel_4,Neel_6}, respectively, larger than that of the other materials in EuM$_2$X$_2$ series \cite{EuZn2P2_Berry2022,EuCd2As2_Inga2011,EuCd2As2_MaJZ_2019_2,Neel_1,Neel_2,Neel_5}.

Using $J_{\parallel}$ and $J_{\perp}$ obtained above, we performed standard Monte Carlo simulations \cite{espins_2022} to calculate the temperature dependence of the heat capacity and magnetization of each material in EuM$_2$X$_2$. A peak is clearly observed in these curves, indicating that the material undergoes a phase transition at this temperature which corresponds to the Néel temperature. Table\ref{table_neel} provides a summary of the Néel temperatures for each material, while Fig.\ref{figneel}(k)(l) illustrate these values graphically. In these figure, black points denotes the experimental data, whereas red and blue points indicate Monte Carlo simulation Néel temperature results derived from heat capacity and susceptibility data, respectively. Notably, the Néel temperatures calculated using a two-parameter Heisenberg model agree very well with the experimental results. For a system with controversial magnetic configurations \cite{EuCd2As2_WangNL2016,EuCd2As2_Rahn2018}, this is an interesting development. This indicates that our Heisenberg model captures the underlying physical nature of the magnetic interactions in the EuM$_2$X$_2$ series, providing a crucial foundation for future research. Meanwhile, we also derived the Néel temperatures of EuCd$_2$Sb$_2$ from the calculation with different Hubbard-$U$ value, which are 13.00 K when $U$ = 3 eV, 7.05 K when $U$ = 5 eV and 3.30 K when $U$ = 7 eV. By comparison, we found the Néel temperature of $U$ = 5 eV is in close agreement with the experimental value of 9.5 K, indicating the choise of $U$ = 5 eV in the above calculation is reasonable. This is further supported by the comparison of optical conductivity presented in the Supplemental Material as illustrated above.

\section{IV. Conclusion}
In summary, we conducted a comprehensive first-principles calculation and magnetic Monte Carlo simulation study on the EuM$_2$X$_2$ family of materials, resulting in two distinct approaches for topological manipulation, while also elucidating the mechanisms of magnetic interactions within the EuM$_2$X$_2$ compounds.
Firstly, we simulated the strength of electron correlation effects in the materials, which can be experimentally controlled through approaches such as applying pressure, by altering the Hubbard-$U$ parameter in the calculation. We found that the strength of electron correlation significantly influences the topological properties of the materials, enabling topological manipulation. Using EuM$_2$X$_2$ with magnetic configuration AFM-Ac as an example, we observed that as the Hubbard-$U$ value increased from 1 eV to 6 eV, the compound underwent multiple topological phase transitions, evolving from trivial insulator to TCI, then to Dirac semimetal, and finally returning to trivial insulator. This provides a promising platform for realizing multiple topological states within a single material.
Secondly, in EuM$_2$X$_2$, as the element X is replaced sequentially from P to As and then to Sb, there is an increase in covalent character, leading to a transition from a positive band gap to a negative one. This alters the sum of parity eigenvalues of valence states and corresponding topological classification. As the weight of element X increases, the EuM$_2$X$_2$ compound transitions from a trivial insulator to a TCI. This represents a second approach to achieving topological manipulation through elemental substitution.
Moreover, using the magnetic exchange coefficients derived from the 'four-state method', we performed Monte Carlo simulations, yielding Néel temperatures that closely match experimental data. This indicates that the model we proposed regarding nearest-neighbor magnetic interactions effectively describe the magnetic interactions within EuM$_2$X$_2$ family of Zintl materials, paving the way for manipulating magnetic structures and potential future applications in magnetism.

\section{Acknowledgment}
We would like to thank Dr.~Z.Y. Liao and Z.Q. Guan of the Institute of Physics, Chinese Academy of Sciences, Dr.~Y. Jiang of Donostia International Physics Center, Paseo Manuel de Lardizábal and Prof.~A. Ptok, Dr.~S. Dan of the Institute of Nuclear Physics, Polish Academy of Science, for helpful discussions. The authors acknowledge the support from the National Key Research and Development Program of China (Grant No. 2022YFA1403800), the National Natural Science Foundation of China (Grants No. 12188101, 11925408), the Chinese Academy of Sciences (Grant No. XDB33000000), and the New Cornerstone Science Foundation through the XPLORER PRIZE.

\bibliography{EuM2X2}

\begin{widetext}
\clearpage
\begin{center}
\textbf{\large Supplemental Material 'Manipulation of topological phase transitions and the mechanism of magnetic interactions in Eu-based Zintl-phase materials'}\\
\vspace{4mm}
{Bo-Xuan Li$^{1,2}$, Ziyin Song$^{1,2}$, Zhong Fang$^{1,2}$, Zhijun Wang$^{1,2}$, Hongming Weng$^{1,2,3,*}$ }\\
\vspace{2mm}
{\em $^1$Beijing National Laboratory for Condensed Matter Physics and Institute of Physics, Chinese Academy of Sciences, Beijing 100190, China\\}
{\em $^2$University of Chinese Academy of Sciences, Beijing 100049, China\\}
{\em $^3$Songshan Lake Materials Laboratory, Dongguan 523808, China}

\end{center}

\setcounter{equation}{0}
\setcounter{figure}{0}
\setcounter{table}{0}
\makeatletter
\renewcommand{\theequation}{S\arabic{equation}}
\renewcommand{\thefigure}{S\arabic{figure}}
\renewcommand{\bibnumfmt}[1]{[S#1]}

This supplemental material contains the orbit projections of  EuCd$_2$As$_2$ with AFM-Aa magnetic configuration FIG.\ref{sfig1} and the comparison between calculated optical conductivity and experimental transmission results of EuZn$_2$As$_2$ with the AFM-Aa magnetic configuration FIG.\ref{sfig2}.

\begin{figure}[h!]
    \centering
    \includegraphics[width=0.7\linewidth]{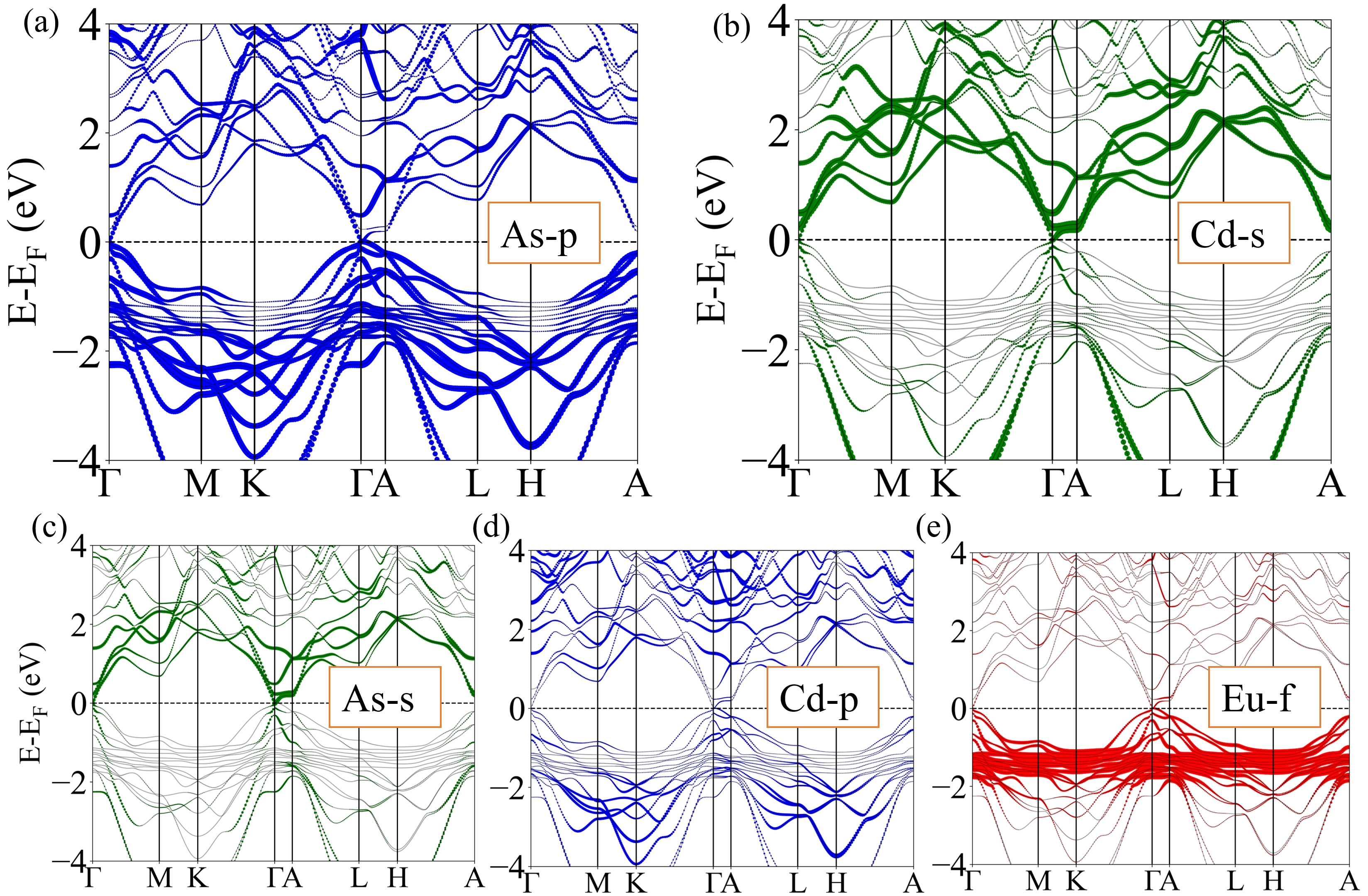}
    \caption{(a)-(e) show different orbit projections of EuCd$_2$As$_2$ with AFM-Aa magnetic configuration}
    \label{sfig1}
\end{figure}

\begin{figure}[h!]
    \centering
    \includegraphics[width=0.7\linewidth]{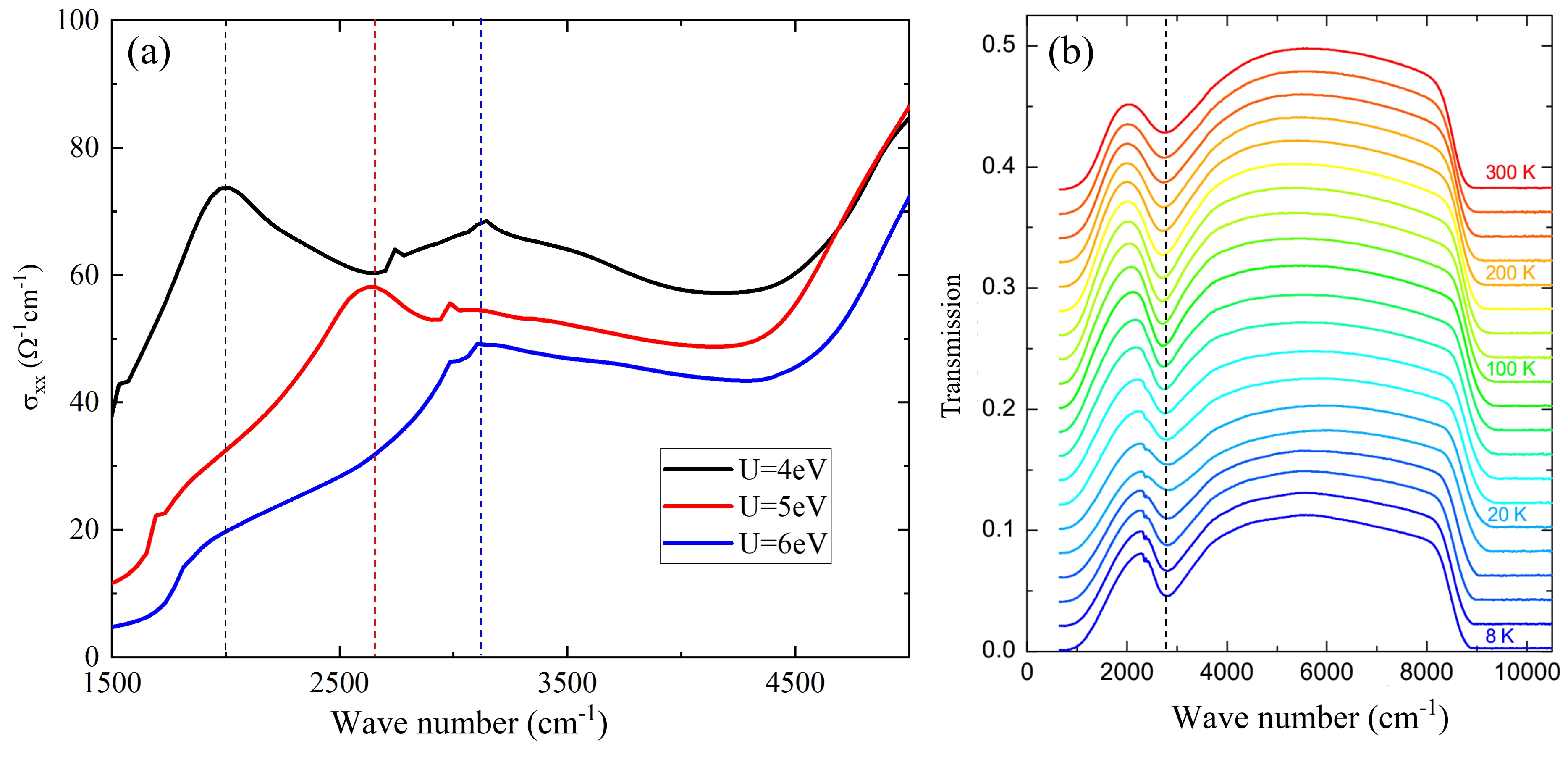}
    \caption{(a) Calculated optical conductivity of EuZn$_2$As$_2$ with the AFM-Aa magnetic configuration for various Hubbard-$U$ values. The dotted lines indicate the peak around the wave number of 3000 cm$^{-1}$. (b) Transmission results of EuZn$_2$As$_2$ with the AFM-Aa magnetic configuration as a function of wave number at different temperatures.}
    \label{sfig2}
\end{figure}

\end{widetext}
\end{document}